\documentclass[pdftex,twocolumn,epjc3]{svjour3}          

\RequirePackage[T1]{fontenc}

\smartqed  

\RequirePackage{graphicx}
\RequirePackage{mathptmx}      
\RequirePackage{flushend}
\RequirePackage[numbers,sort&compress]{natbib}
\RequirePackage[colorlinks,citecolor=blue,urlcolor=blue,linkcolor=blue]{hyperref}

\journalname{Eur. Phys. J. C}

\newcommand{\be}{\begin{equation}}
\newcommand{\ee}{\end{equation}}
\newcommand{\bea}{\begin{eqnarray}}
\newcommand{\eea}{\end{eqnarray}}
\newcommand{\ii}{\mathrm{i}}

\usepackage{amsmath}
\usepackage{graphicx}
\usepackage{subfigure}

\begin{document}

\title{ Emergent spontaneous symmetry breaking and emergent symmetry restoration in rippling gravitational background. }

\author{Maxim A.~Kurkov\thanksref{e1,addr1}
       }

\thankstext{e1}{e-mail: max.kurkov@gmail.com}

\institute{CMCC-Universidade Federal do ABC, Santo Andr\'e, S.P., Brazil\label{addr1}
}

\date{Received: date / Accepted: date}

\maketitle

\begin{abstract}
We study  effects of a rippling gravitational background on a scalar field with a double well potential, 
 focusing on the analogy with the well known dynamics of the Kapitza's pendulum.
The  ripples are rendered as infinitesimal but rapidly oscillating perturbations of the scale
factor. We find that the resulting dynamics crucially depends on a value of the parameter $\xi$ in the $\xi \,R\, \phi^2$ vertex. 
For the time-dependent perturbations of a proper form
the resulting effective action is generally covariant, and  at a high enough frequency at $\xi<0$ and at $\xi>1/6$  the  effective potential 
has a single minimum at zero, thereby restoring spontaneously broken symmetry of the ground state. 
On the other side, at $0<\xi< 1/6$   spontaneous symmetry breaking emerges even when it is absent in the unperturbed  case. 
\end{abstract}

\section{Introduction}
Oscillating gravitational backgrounds attract attention of theoretical physicists  in  various contexts
starting from  quantum decoherence \cite{Power:1998wh, Wang:2006vh}
and finishing some cosmological \cite{Durrer:1995mz} and extra dimensional constructions 
\cite{Collins:2001ic,Collins:2001mz,Collins:2001ni}. Effects of a rippling scale factor
on a \emph{minimally} coupled quantized scalar field were studied in \cite{Hwang:1993cv}.
One can also find some interesting consequences of rippling parameters in \cite{Mangano:2015pha,deCesare:2016dnp}.
The first famous example of a rippling background, however, is not related to
quantum physics or cosmology - it is the famous Kapitza's pendulum \cite{Kapitza1,Kapitza2}, i.e. a classical pendulum whose
 suspension point rapidly oscillates with a small amplitude. The highly nontrivial and counterintuitive classical dynamics of this simple object inspired us to write this article.  As we show below already at classical level small but rapidly oscillating perturbations of the scale factor may  affect the ground states of physically relevant models  in a quite interesting way.

In this paper we study the effect 
of the rapidly oscillating gravitational field  $g_{\mu\nu}$ on a classical dynamics of the scalar field $\phi$.
The latter is described by the most general action\footnote{Hereafter the lower index specifies which metric tensor is used
in order to construct a given quantity. }
\be
\small{S_g\left[\phi\right] = \int d^4 x\sqrt{-g}\, \left\{\frac{1}{2}\phi\left(-\nabla^2_g -\xi R_{g} + \mu^2\right)\phi   - \frac{1}{4!}\lambda \phi^4 \right\}, }\label{Sint}
\ee 
which leads to renormalizable\footnote{This statement refers to quantum field theory on a \emph{classical} gravitational background.} quantum field theory in four dimensions, which is invariant upon the reflection $\phi \longrightarrow -\phi$.  Such an action is of special importance, since the Higgs field of the Standard Model is described by the action of such a form.
We consider the  metric tensor given by the product of the slowly varying bare\footnote{In what follows the word ``bare" is addressed to the unperturbed theory i.e. at $\omega=0$.} metric  $\bar{g}_{\mu\nu}$ and the highly oscillating conformal factor, which is very close to one:
\be 
g_{\mu\nu} = \bar{g}_{\mu\nu}e^{2\alpha\cos{(\omega t)}}, \label{gdef}
\ee 
where the dimensionless amplitude $\alpha$ is much smaller than $1$, the frequency
$\omega$ is much larger than $\mu$ and
invariants of the metric tensor $\bar g_{\mu\nu}$ (and its derivatives) of the dimension [${\mbox{length}}^{-1}$]. The main goal of the present paper is to
generalize the result of Kapitza to this model. In particular, we show that solutions of 
the equation of motion, which comes out from the action \eqref{Sint}, split into two parts
\be
\phi = \bar\phi  + \delta\phi,
\ee
where $\delta \phi$ is a rapidly oscillating and small correction, which approaches to zero as $\omega^{-1}$ in the high frequency limit, while the dynamics of the
slowly varying field $\bar\phi$ is described by the classical action, which we establish. 
Surely due to a manifest frame dependence of the perturbation law \eqref{gdef}, the effective action 
generally speaking depends on the choice of the coordinates, which are used to perturb the bare metric tensor $\bar g_{\mu\nu}$.

The most interesting
result is related to the situation, when the oscillating metric takes the simple form  \eqref{gdef}  in the 
comoving frame\footnote{Actually the result holds for a bigger class of perturbations, see the discussion in Sec. 3.}.
In such a case the effective action has exactly the same structure  \eqref{Sint}, where the quantities $g_{\mu\nu}$ and $\phi$
are replaced correspondingly by $\bar g_{\mu\nu}$ and $\bar\phi$, 
\be
\small{S^{\rm eff}_{\bar g}\left[\bar\phi\right] = \int d^4 x\sqrt{-\bar g}\, \left\{\frac{1}{2}\bar\phi\left(-\nabla^2_{\bar g} -\xi R_{\bar g} + \mu^2_{\rm eff}\right)\bar\phi   
- \frac{1}{4!}\lambda \bar\phi^4 \right\},} \label{Seff}
\ee
therefore it is generally covariant, 
and the only effect of the oscillations
is an additive renormalization of the mass parameter $\mu^2$.  A sign of 
the correction $\Delta\mu^2\equiv \mu_{\rm eff}^2 - \mu^2$ nontrivially depends on a choice of the coupling
constant $\xi$, and this is probably the most interesting result of the paper. 

We find that at $\xi < 0$ and at $\xi > 1/6$ the correction $\Delta\mu^2$ is always negative, and at the high enough
frequency $\omega$ it can easily compensate the bare mass parameter, what quite naturally implies that the effect of the rapid oscillations
is similar to an increase in temperature: the effective potential (in contrast to the original one) has just a single minimum at $\phi = 0$,
and spontaneous symmetry breaking of the bare   theory vanishes. All particles which obtain their mass due
to the Higgs mechanism become massless.
 
In the another  regime $0< \xi < 1/6$ the situation changes qualitatively. The correction $\Delta\mu^2\equiv \mu_{\rm eff}^2 - \mu^2$ becomes positive, 
and even when the mass parameter is absent in the bare  theory\footnote{or it is negative}, it appears
in the effective one. Such an effect can be regarded as  some sort of \emph{emergent} spontaneous symmetry breaking.  At $\xi = 1/6$ (naturally) and $\xi = 0$ (less naturally) the effect is absent. 

An influence of  time dependent (but not oscillating) gravitational backgrounds on the vacuum state of a scalar field was  studied in the context of inflation in \cite{Vilenkin:1982wt}, and the result crucially depends on the non minimal coupling $\xi$ as well,
in particular at $\xi = 1/6$ the influence is minimal.
 Nevertheless, the effects discussed 
in \cite{Vilenkin:1982wt} and here are different, since in contrast to our setup in the framework of \cite{Vilenkin:1982wt} at $\xi=0$ the vacuum expectation value 
grows up exponentially with time. 

This paper is organized as follows. In Sec.~2 we briefly describe relevant aspects of the dynamics of the Kapitza's pendulum in order
to apply the same technique in the forthcoming discussion. In Sec.~3 we derive the effective action \eqref{Sint} and establish
the renormalized mass parameter $\mu_{\rm eff}^2$. Sec.~4 is devoted to numerical illustrations of the results obtained for the flat  bare metric $\bar g_{\mu\nu}$.

\section{Kapitza's pendulum: a brief technical review}
The mathematical pendulum is described by the following equation of motion
\be
 \ddot{f} +  g\sin \left( f \right) = \ddot{f} +  g f +\mathcal{O}\left(f^3\right) =0, \label{pendorig}
\ee
where $f(t)$ is the angular coordinate (we consider the length to be equal to one), and $g>0$ is the free-fall acceleration;
hereafter the dot stands for differentiation with respect to time.
The only stable equilibrium position is located at $f = 0$, but the situation changes dramatically,
when the pivot  vibrates with the high frequency $\omega$. In such a case the equation of motion reads
\be
\ddot{f}  + \left( g+\alpha{\omega}^{2}
\cos \left( \omega\,t \right)  \right) \sin \left( f
 \right) =0, \label{pendoscil}
\ee
where the constant $\alpha$ stands for the amplitude of the oscillations of the suspension, and it is supposed to be small.  The most amazing fact is the appearance of a new stable equilibrium point in the inverted position 
i.e. at $f = \pi$. More precisely
the solution of the equation of motion is given by a sum of the two functions $\bar f(t)$ and $\delta f (t)$, where the former
satisfies some ``effective" equation of motion, which in particular exhibits slow oscillations in the vicinity of $f = \pi$, when the frequency  $\omega$ is high enough. The latter
is a rapidly oscillating correction of the order\footnote{Actually this asymptotic takes place in the double limit 
$\alpha\rightarrow 0$,
and $\omega\rightarrow \infty$, while their product $\alpha\cdot\omega \equiv \gamma$ is a constant.} 
of $\mathcal{O}\left(\omega^{-1}\right)$.
It is remarkable  that  one can relatively easily  give a precise mathematical description of such a nontrivial and unexpected phenomenon. Below we present a derivation of this
well known result in the form which is the most suitable and clear for the forthcoming analysis
of the scalar field in a rapidly oscillating gravitational background. \\

\noindent {\bf Step 1. Proper ansatz.}\\
A numerical analysis of the equation \eqref{pendoscil} hints us, that its solution splits into the slowly varying part $\bar f(t)$ and the rapidly oscillating but small correction $\delta f$. It is natural to expect that the correction $\delta f$  oscillates with the same frequency $\omega$, therefore let us look for the solution in a form of an asymptotic expansion in inverse powers of $\omega$ 
\be
f = \bar f + \frac{1}{\omega}\left(A \sin(\omega t) + B \cos(\omega t) \right)+ \mathcal{O}\left(\frac{1}{\omega^2}\right), \label{ans}
\ee
where the (yet undetermined)  functions $A(t)$ and $B(t)$ vary slowly\footnote{At this point this is just an
 assumption to be  checked aposteriori, after all the functions are found.}. 
Substituting the ansatz \eqref{ans} and $\alpha = \gamma\cdot \omega^{-1}$ in the equation of motion \eqref{pendoscil}
  we obtain at the leading order at large $\omega$:
 \bea
0 &=& \mathcal{F}(t) = \omega \left\{ \sin \left( \bar f \left( t \right)  \right) \gamma\,\cos \left( \omega
\,t \right) \right. \nonumber \\
 &-& \left. B \left( t \right) \cos \left( \omega\,t \right) -A
 \left( t \right) \sin \left( \omega\,t \right)\right\} + \mathcal{O}(\omega^0), \label{asex}
 \eea
what defines $A(t) = 0 $ and $B(t) = \gamma \sin \left( \bar f  \right)$. From this moment on
the ansatz \eqref{ans} contains just one undetermined function $\bar f$. \\

\noindent {\bf Step  2.  Averaging over rapid oscillations.}\\
Substituting $A(t)=0$ and $B(t) = \gamma \sin \left( \bar f  \right)$ in the next to the leading order of our asymptotic expansion \eqref{asex}
we immediately arrive to the equation which determines $\bar f(t)$.
\bea
0 = \mathcal{F}(t) = \ddot{\bar f} +\sin \left( \bar f  \right) g-2\,\cos \left( \bar f 
 \right)  \dot{\bar f}\gamma
\,\sin \left( \omega\,t \right)\nonumber \\ 
+\cos \left( \bar f 
 \right) {\gamma}^{2} \left( \cos \left( \omega\,t \right)  \right) ^{
2}\sin \left( \bar f  \right)  + \mathcal{O}(\omega^{-1}) \label{interm}
\eea
Now let us average the function $\mathcal{F}(t)$ over the period of rapid oscillations $\frac{2\pi}{\omega}$: 
\be
\langle \mathcal{F}\rangle(t) \equiv \frac{\omega}{2\pi} \int_{0}^{\frac{2\pi}{\omega}} d\tau \mathcal{F}(\tau + t). \label{intinterm}
\ee
Since by our assumption the quantity $\bar f$ and its derivative vary slowly during the period of rapid oscillations, 
one can easily compute the integral \eqref{intinterm} in the limit $\omega\rightarrow\infty$, and the answer reads: 
\be
   \ddot{\bar f} +\sin \left( \bar f \right) g+\frac{1}{2}\,\cos \left( \bar f 
 \right) {\gamma}^{2}\sin \left( \bar f  \right) = 0, \quad \gamma \equiv \omega \cdot\alpha \ll \omega. \label{finalKapEq}
\ee
The equation \eqref{finalKapEq} is autonomous and it does not contain large parameters anymore, therefore $\bar f$ is indeed slowly varying function\footnote{Unless one chooses large initial velocity.}
, 
what justifies 
 all our assumptions.
 
Small deviations of $\bar f$ from zero satisfy the standard pendulum equation \eqref{pendorig}
where g is replaced by $g_{\rm eff}$
\be
g^{\rm down}_{\rm eff} = g+\frac{1}{2} \alpha^2\omega^2, \label{gdown}
\ee
thus in the vicinity of this equilibrium point one has an additive (finite) renormalization of the free fall
acceleration constant $g$.
On the other side, and this is even more interesting, small deviations from $\bar f = \pi$ satisfy linearized version of Eq.~\eqref{pendorig}
with 
\be
g_{\rm eff}^{\rm up} = -g+\frac{1}{2} \alpha^2\omega^2 > 0,\quad\mbox{at} \quad |\omega\cdot\alpha| > {\sqrt{2g}}, \label{gup}
\ee 
therefore in this regime the pendulum
starts to oscillate near its upper position  with the slow frequency 
$\Omega\equiv\sqrt{ -g+\frac{1}{2} \alpha^2\omega^2}$ $\ll$ $\omega$. 
Below we consider the system described by the action \eqref{Sint} along similar lines, and we find that  it shares some similarities with the Kapitza's pendulum.

\section{Scalar field in oscillating gravitational background}
Now we study the effect of the rapid oscillations defined by \eqref{gdef} on the dynamics of the scalar field $\phi$,
which obeys the classical equation of motion
\be
-\frac{1}{\sqrt{-g}}\frac{\delta S_g\left[\phi\right]}{\delta \phi} =  \left(\nabla^2_{ g} +\xi R_{ g} - \mu^2\right)\phi   
+ \frac{1}{3!}\lambda \phi^3  = 0. \label{geneq}
\ee
The gravitational background is considered to be given, and we assume that the back-reaction
of the scalar field on its dynamics is negligibly small. 
We simplify the forthcoming analysis  substituting
\be
\phi = e^{-\alpha \cos(\omega t)}\tilde\phi. \label{change}
\ee
On the one hand,  due to the local Weyl
invariance of the action \eqref{Sint} at $\xi=\frac{1}{6}$ the equation \eqref{geneq}  rewritten in terms of $\tilde\phi$
has simpler dependence on $\alpha$ and $\omega$. On the other hand,   we are interested in the evolution of the slowly varying part of $\phi$, which coincides for $\tilde\phi$ and $\phi$.
 
In order to carry out the computations 
it makes sense to rewrite the equation of motion splitting explicitly rapidly and slowly varying
quontities  
\be
\left(\bar g^{\mu\nu}\partial^2_{\mu\nu} + Y^{\mu}\partial_{\mu} +\xi R_{\bar g}  - \mu^2\right)\tilde\phi + \frac{\lambda}{3!}\tilde\phi^3 + f_{\alpha} = 0, \label{geneq2}
\ee 
where $\bar g^{\mu\nu}$, 
$Y^{\mu}\equiv \frac{1}{\sqrt{-\bar g}}\left(\partial_{\nu}\bar g^{\mu\nu}\sqrt{-\bar g}\right)$ and $ R_{\bar g}$
vary slowly by construction. The function $f_{\alpha}$ absorbs all influence of the rapidly oscillating exponential
factor in \eqref{gdef}, and it is defined by
\bea
 f_{\alpha} &\equiv&\left[ \left(6\xi - 1\right)\left(-\omega^2\alpha{\,\bar g}^{00}\left(\cos(\omega t)-\alpha\sin^2(\omega t) \right) \nonumber \right. \right.\\
 &-&  \left.\left.\omega\alpha Y^{0}\sin(\omega t) \right) 
+\mu^2\left(1 - e^{2\alpha\cos(\omega t)}\right)\right]\tilde\phi \label{falpha}.
\eea
Below we study the leading asymptotic at large $\omega$ and small $\alpha$
keeping  $\alpha = \gamma/\omega = \mathcal{O}(\omega^{-1})$ by analogy with the Kapitza's pendulum.
\\

\noindent{\bf Step 1. Proper ansatz.} \\
Inspired by the example discussed in the previous section we look for the solution in a form of the following 
asymptotic ansatz:
\be
\tilde\phi = \bar{\phi} + \frac{1}{\omega}\left(T\cos(\omega t) + W\sin(\omega t)\right) 
+\mathcal{O}\left(\frac{1}{\omega^2}\right), \label{ans2}
\ee
where $T$ and $W$ are slowly varying functions of all coordinates.
 We do not write the tilde over $\bar\phi$, since, as we said above, in the limit of infinitely large $\omega$
(and infinitely small $\alpha$) the quantities $\phi$ and $\tilde\phi$ coincide. 
Substituting the ansatz \eqref{ans2} in the equation of motion \eqref{geneq2} we immediately
determine $T$ and $W$:
\bea
0=\mathcal{F}&\equiv& \omega\left(-6\,\xi\,\gamma\,{\bar g}^{00}\cos \left( \omega\,t \right) \bar{\phi}-{\bar g}^{00}
T\cos \left( \omega\,t \right) \right. \label{asexp2}\\
 &-& \left. {\bar g}^{00}W\sin \left( \omega\,t
 \right) +\gamma\,{\bar g}^{00}\cos \left( \omega\,t \right) \bar{\phi}
\right) + \mathcal{O}\left(\omega^0\right),  \nonumber
\eea
what implies
\be
T = (1-6\xi)\gamma\bar{\phi}, \quad W = 0. \label{TW}
\ee
As in the ``canonical" example there is just one undetermined function $\bar{\phi}$.
\\

\noindent{\bf Step 2. Averaging over rapid oscillations.} \\
Exactly as we did in the previous section, substituting \eqref{TW} in the next to the leading
order of our asymptotic expansion \eqref{asexp2} we derive the equation which determines $\bar{\phi}$.
\bea
0 &=&\mathcal{F}\equiv \left({\bar g}^{\mu\nu}\partial^2_{\mu\nu} + Y^{\mu}\partial_{\mu} 
+\xi R_{\bar g}  - \mu^2\right)\bar{\phi} 
+ \frac{\lambda}{3!}\bar{\phi}^3 \nonumber \\
&+& (6\xi-1)\left[2\gamma \,{\bar g}^{\mu 0}\left(\partial_{\mu} \bar{\phi}\right) \sin(\omega t) \right. \\
& 
+& \left. {\gamma}^{2}\,{\bar g}^{{00}}\,\bar{\phi}\, \left( 6\,
 \left( \cos \left( \omega\,t \right)  \right) ^{2}\xi-\cos(2\omega t) \right)\right] + \mathcal{O}\left(\frac{1}{\omega}\right) \nonumber
\eea
Averaging over the period of rapid oscillations (c.f. Eq. \eqref{intinterm}) and passing to the limit 
of large $\omega$ we obtain the final effective equation for $\bar\phi$:
\bea
0 &=&\lim_{\omega\rightarrow\infty} \langle \mathcal{F} \rangle
= \left(\nabla^2_{ \bar g} +\xi R_{ \bar g} - \mu^2 + 3\,{\gamma}^{2}{\bar g}^{{00}}\,\xi\, \left( 6\,\xi-1 \right)\right)\bar\phi \nonumber\\
   &+& \frac{1}{3!}\lambda \bar\phi^3  . \label{semifinal}
\eea

The structure of Eq.~\eqref{semifinal} immediately suggests us,  which oscillations of the form \eqref{gdef}
are the most interesting. Surely we wish to obtain a generally covariant expression by the end of the day,
what happens if and only if the simple perturbation equation \eqref{gdef} is written in the frame where ${\bar g}^{{00}}$
does not depend on coordinates e.g. the comoving frame. 
  We emphasize that we are talking about preferred ways of perturbation of the bare metric $\bar g_{\mu\nu}$ rather than preferred coordinate systems.
Starting from now we assume  that the perturbations are ``good" in the mentioned above sense, and\footnote{In principle $\bar g^{00}$ must not be necessary positive and it can be either negative or zero (e.g. the light front coordinates). In the former case 
the effect changes its sign, and in the latter case it is absent.}  
the coordinates in Eq.~\eqref{gdef} are chosen in such a way that $\bar g^{00} = +1$,
so we arrive to the manifestly generally covariant equation 
\be
0 = \left(\nabla^2_{ \bar g} +\xi R_{ \bar g} - \mu_{\rm eff}^2 \right)\bar\phi   
+ \frac{1}{3!}\lambda \bar\phi^3  = - \frac{1}{\sqrt{-\bar g}} \frac{\delta S^{\rm eff}_{\bar g}\left[\bar\phi\right]}{\delta\bar\phi}, \label{final}
\ee
which holds for an arbitrary coordinate system\footnote{Surely, for an arbitrary frame
one has to rewrite Eq.~\eqref{gdef}, which defines the perturbations,  according to the tensor transformation law.}, and
 has exactly the same structure as Eq.~\eqref{geneq}. The effective mass parameter is given by
\be
\mu_{\rm eff}^2 = \mu^2 + 3\,\alpha^2 \omega^2 \xi\, \left( 1-6\,\xi \right) \label{muren} \,\ll \, \omega^2.
\ee
As we announced in the introduction, this quantity can be either positive or negative depending on the
ratios between the parameters, what may dramatically affect the spontaneous symmetry
breaking. The particular form \eqref{muren} of the additive renormalization of the mass parameter $\mu^2$ suggests
us to consider two different regimes of the non minimal scalar-tensor coupling, defined
by the dimensionless coupling constant $\xi$.

\begin{itemize}
\item{At $\xi < 0$ and $\xi>1/6$ the correction to the mass parameter is negative, and
at 
\be
|\omega \cdot\alpha|  > \frac{\mu}{\sqrt{3\xi(6\xi - 1)}}, \quad \mu^2 > 0,  \label{cond1}
\ee
the effective mass parameter becomes negative, what restores a spontaneously broken symmetry of the ground state. This effect is similar to the dynamical stabilization of the upper equilibrium  position  of the Kapitza's pendulum, c.f. Eq.~\eqref{gup}.
}
\item{At $0<\xi<1/6$ the correction is positive, what is similar to the effective renormalization
of the free fall acceleration in the context of the Kapitza's pendulum oscillating near its lower equilibrium, c.f. \eqref{gdown}. This regime
becomes more interesting when the ``bare theory" is not spontaneously broken, i.e. at $\mu^2 = -m^2  <0$. 
In such a situation at
\be
\quad\quad\quad |\omega \cdot \alpha|  > \frac{m}{\sqrt{3\xi(1 - 6\xi)}},\quad m^2 \equiv -\mu^2 > 0,  \label{cond2}
\ee
the effective mass parameter $\mu^2_{\rm eff}$ becomes positive in contrast to the bare one $\mu^2$,
what implies emergent spontaneous symmetry breaking. One can easily see, that this effect reaches its
maximum at $\xi = 1/12$.
}
\end{itemize}
If for some application one prefers to avoid the mentioned above effects, one has to work with the minimal $\xi = 0$ or the conformal $\xi = 1/6$ couplings.  
  \\

\noindent{{\it{\bf Remark}} We notice that the coordinate $x_0\equiv t$ must not be necessary the conformal
time, but it can be also the comoving time $t_c$. Let us fix the gauge by the following condition 
\be
\bar g_{00} = 1,\quad \bar g_{0j} = 0, \quad j = 1,2,3, \label{comfr}
\ee 
so the square of the infinitesimal four interval reads (c.f. Eq.~\eqref{gdef}):
\be 
ds^2 = \left(dt^2 -  d\vec{x}^2\right)e^{2\alpha\cos{(\omega\,t})} = dt_c^2 - \left(d\vec{x}^2\right)e^{2\alpha\cos{(\omega\,t_c})},
\ee
where $d\vec{x}^2\equiv\bar g_{ij}dx^i dx^j$, $i,j = 1,2,3$. In other words instead of consideration of the fluctuations of the form \eqref{gdef}
we could have considered 
\be
g_{00}  = \bar g_{00},\quad g_{ij} = \bar g_{ij} e^{2\alpha\cos(\omega\, t)}, \quad i,j = 1,2,3 \label{altern}
\ee
from the very beginning\footnote{upon the condition \eqref{comfr}}, what is more natural for  Friedmann-Lema\^\i{}tre-Robertson-Walker metric.
All the results viz the final effective equation \eqref{final}, the renormalization of the mass paremeter \eqref{muren}
and the conditions \eqref{cond1},  \eqref{cond2}, which define the phase transitions, remain absolutely the same.
Note that in \eqref{altern} we wrote $t$ instead of $t_c$, since up to a correction of the order of $\mathcal{O}\left(\omega^{-1}\right)$ 
these two quantities (but not their derivatives!) coincide:
\be
t = \int^{t_c}_0 dz e^{-2\alpha\cos(\omega z)} = t_c +\mathcal{O}\left(\omega^{-1}\right),
\quad \alpha \equiv \frac{\gamma}{\omega} \ll1.
\ee
}

So far we discussed the ripples along the time direction $t\equiv x^0$. One can easily elaborate the oscillations
along some spatial direction e.g. $z\equiv x^3$ in a similar manner. Replacing $t$ by $z$ in the oscillating
conformal factor in \eqref{gdef} and repeating all the discussion we arrive to \eqref{semifinal}, where $\bar g^{00}$ is replaced by $\bar g^{33}$.
The effective equation becomes generally covariant if and only if the (modified) perturbation law \eqref{gdef} is written
in the  coordinate system, where $\bar g^{33} = \mbox{const}$, for example in the Gaussian normal coordinates
in respect to the hypersurface of constant $z$.  It is worth noting that  for such a choice of coordinates in the perturbation law
the effect of this ``spatial" rippling has the \emph{opposite} sign with respect to the time-dependent rippling introduced in the comoving frame,
since $\bar g^{33} = -1$ in the Gaussian normal coordinates.

In the next section we illustrate numerically how this mechanism works on a few examples.

\section{Numerical illustrations}
As we have seen in the previous section the behavior of the scalar field in a rapidly oscillating
gravitational background crucially depends on two entries: the parameter $\gamma\equiv \alpha\cdot\omega$,
which comprises the information on the perturbation, and  the parameter $\xi$,
which is responsible for the non minimal tensor-scalar coupling.

\begin{figure*}[p]
\centering
\includegraphics[scale = 0.6]{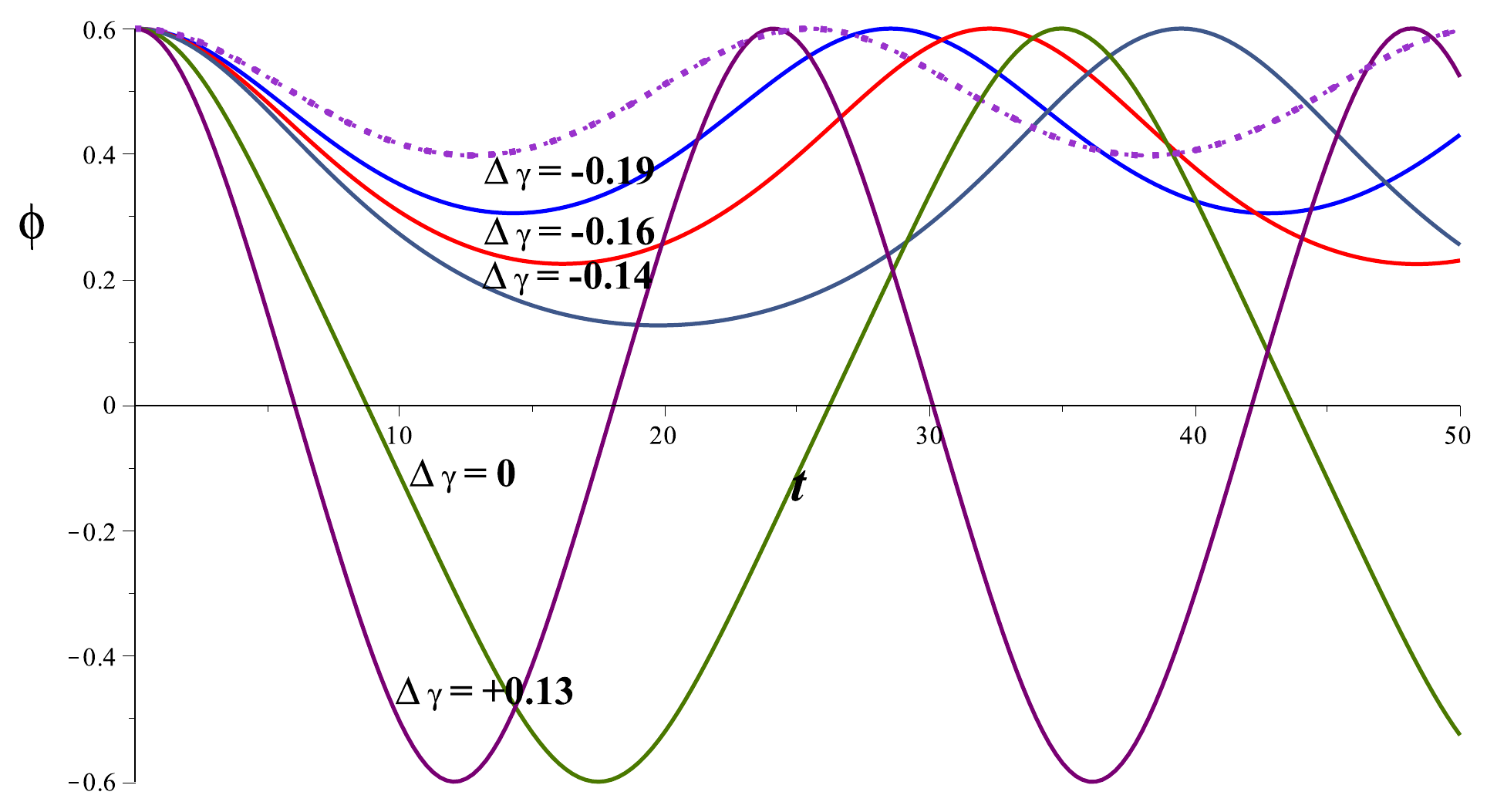}
\caption{\label{fig1}\emph{ Emergent symmetry restoration, $\gamma$ dependence.} 
\small{ 
The solid lines
represent $\phi(t)$ for various choices of the parameter $\gamma$ in the vicinity
of its critical value $\gamma_0$.  Above each solid curve the deviation $\Delta \gamma\equiv \gamma-\gamma_0$ 
from the critical value is specified. The dotted line represents the solution of the bare equation of motion.
 The parameters  are chosen as follows: $\xi = -\frac{1}{12}$, $\mu = 0.18$,  hence $\gamma_0\simeq 0.294$; $\lambda = 0.75$,
$\omega = 2500$; for each curve  $\alpha = (\gamma_0 +\Delta\gamma)/\omega $. The initial conditions are chosen as follows: $\phi(0) = 0.6$, $\dot\phi(0) = 0$.} }
\end{figure*}
\begin{figure*}[p]
\centering
\subfigure[]{\includegraphics[scale = 0.28]{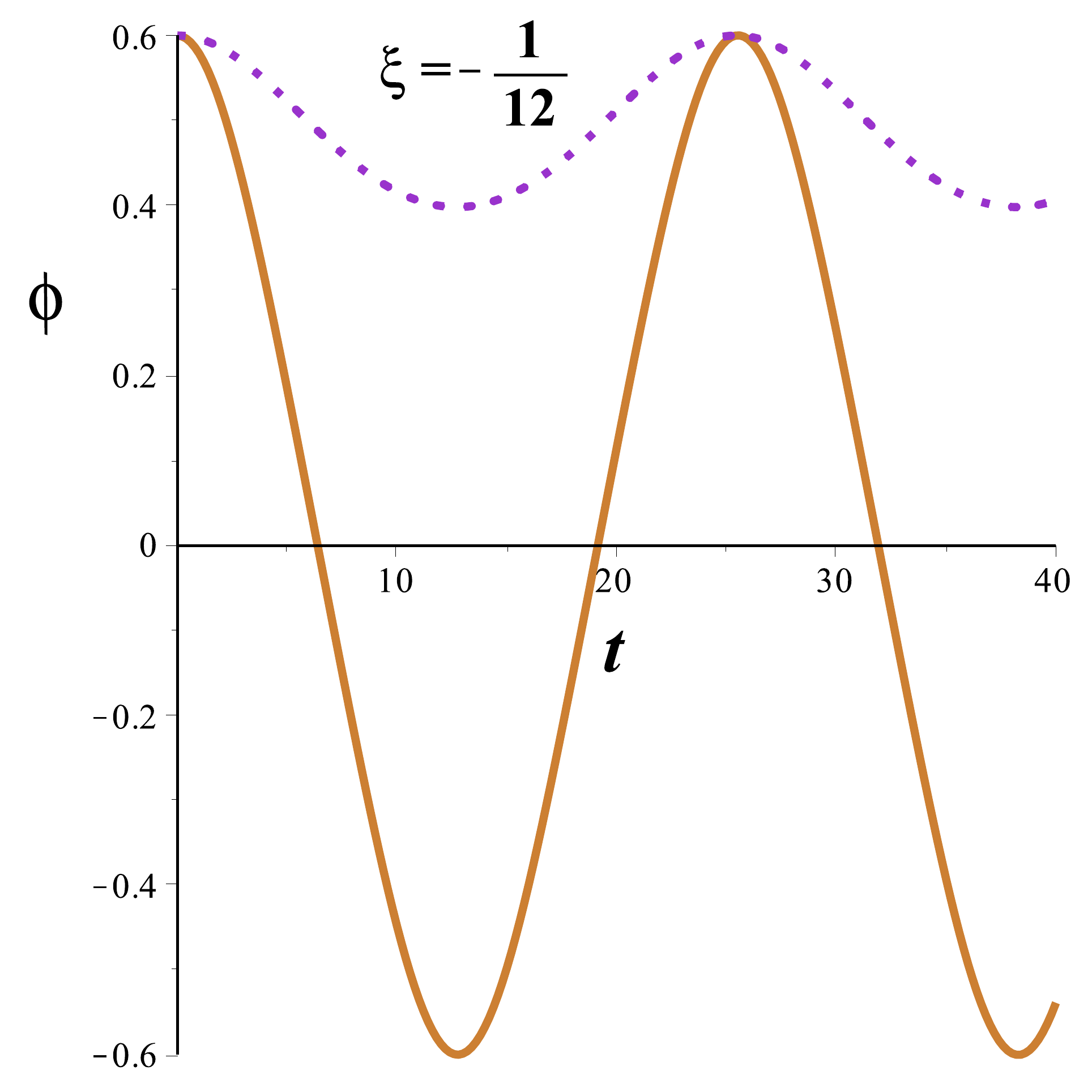} \label{fig2a}  }
\subfigure[]{\includegraphics[scale = 0.28]{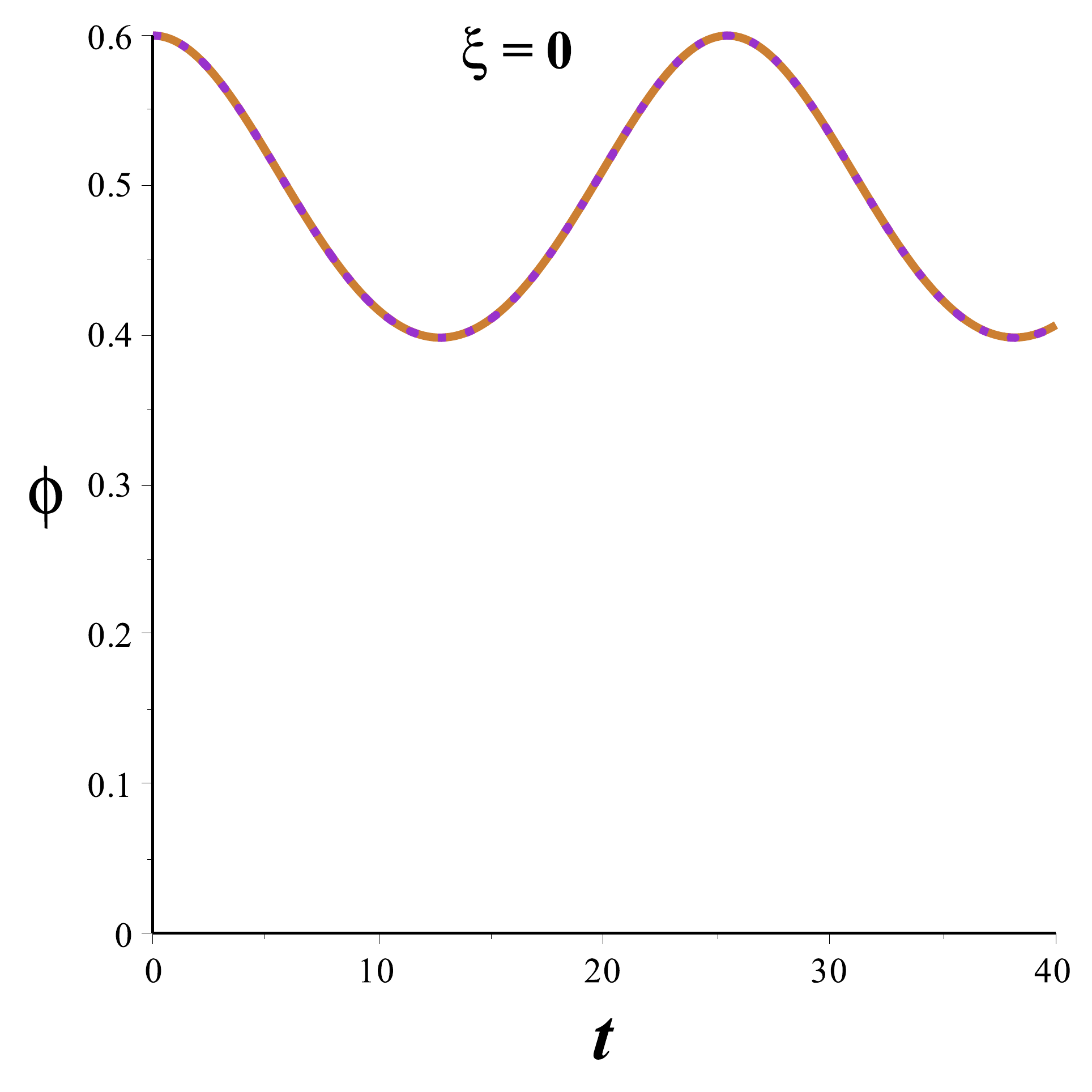}  \label{fig2b} }
\subfigure[]{\includegraphics[scale = 0.28]{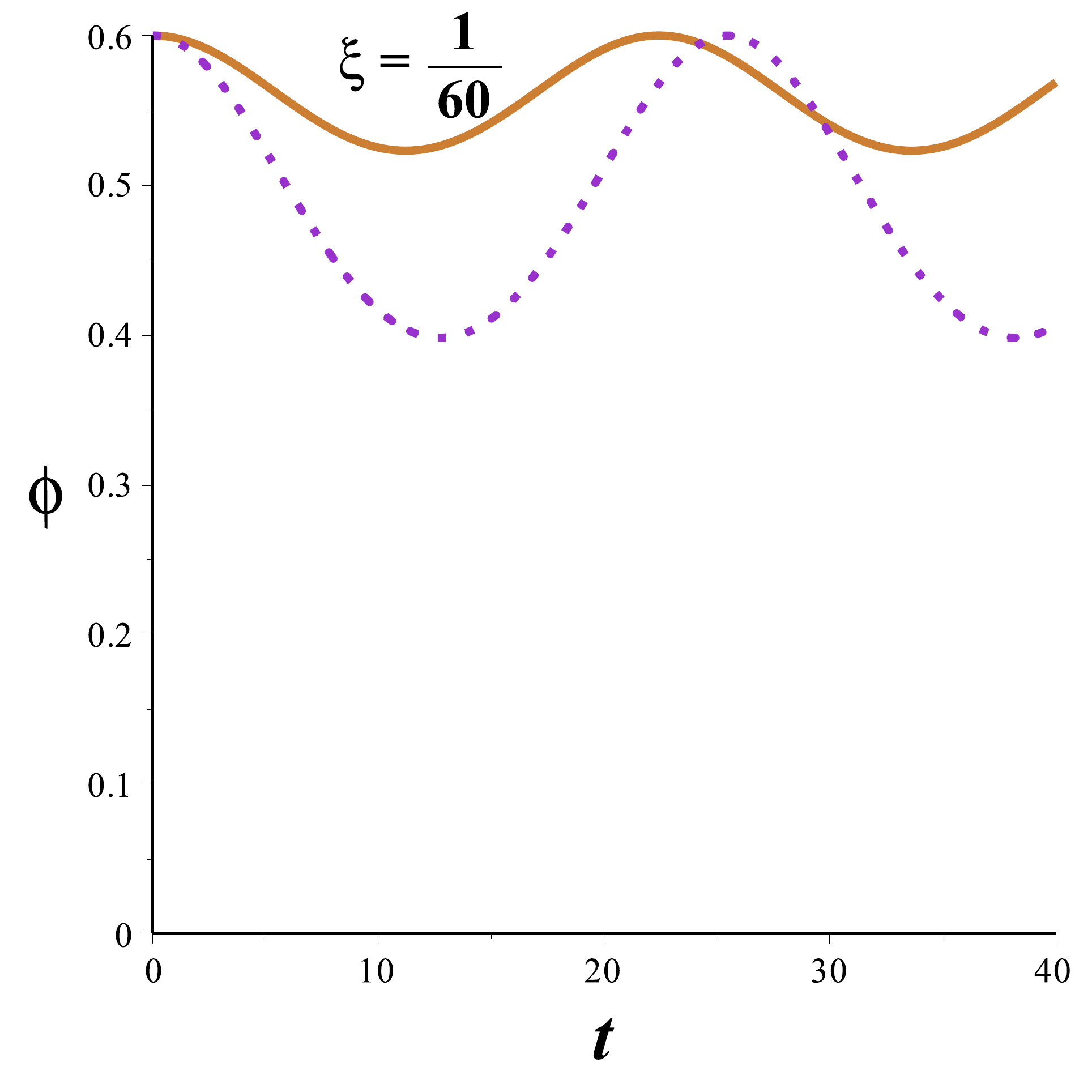} \label{fig2c}  }
\subfigure[]{\includegraphics[scale = 0.28]{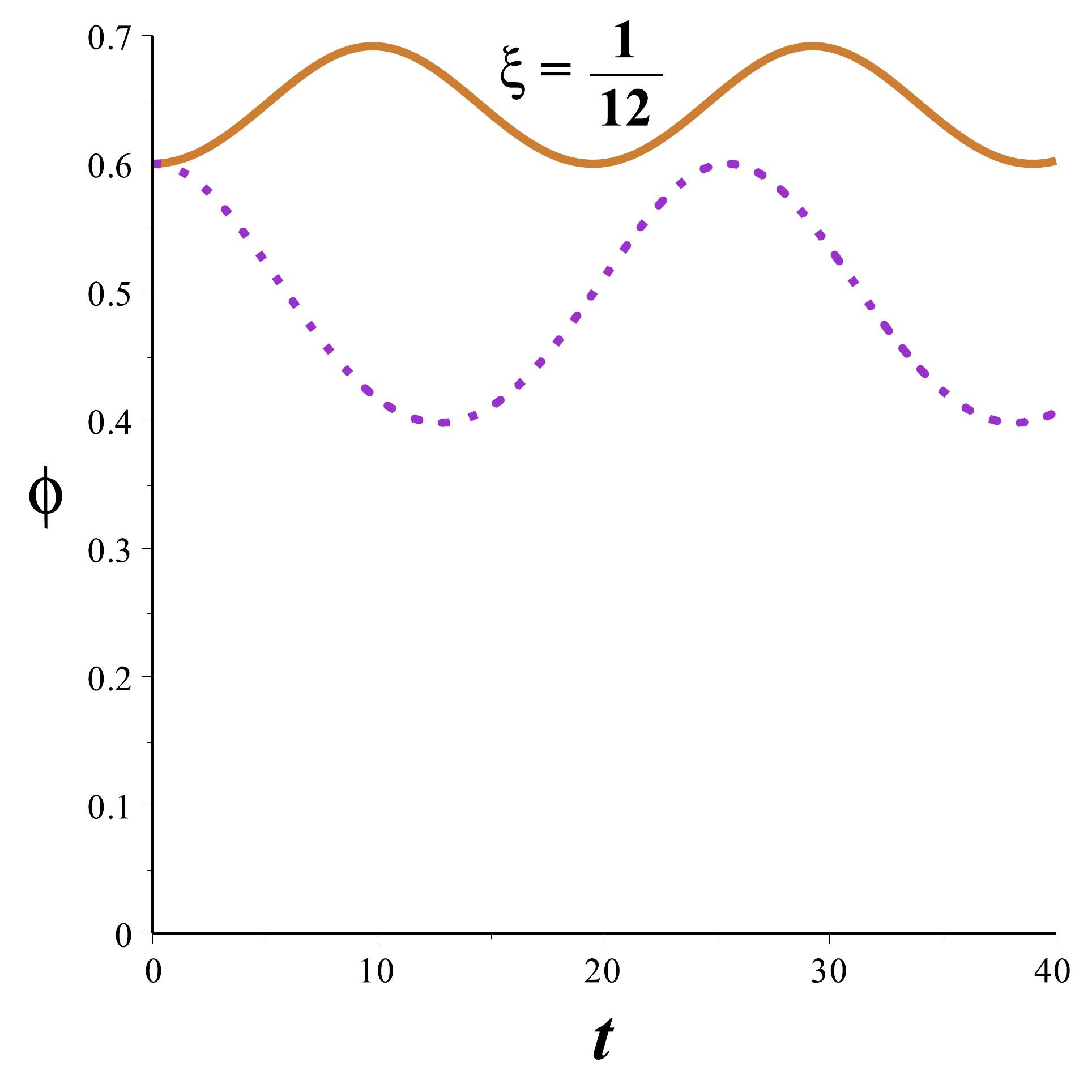} \label{fig2d} }
\subfigure[]{\includegraphics[scale = 0.28]{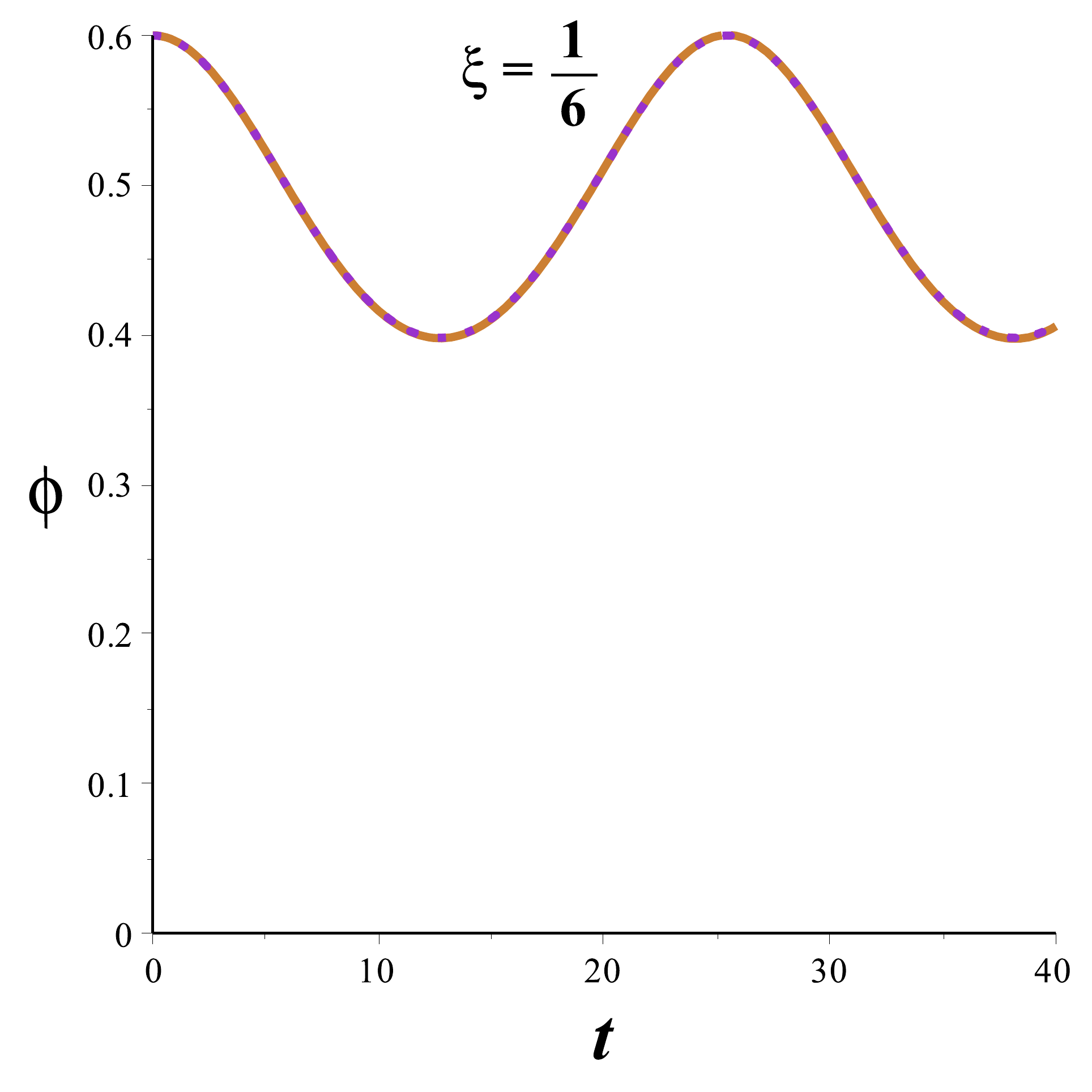} \label{fig2e} }
\subfigure[]{\includegraphics[scale = 0.28]{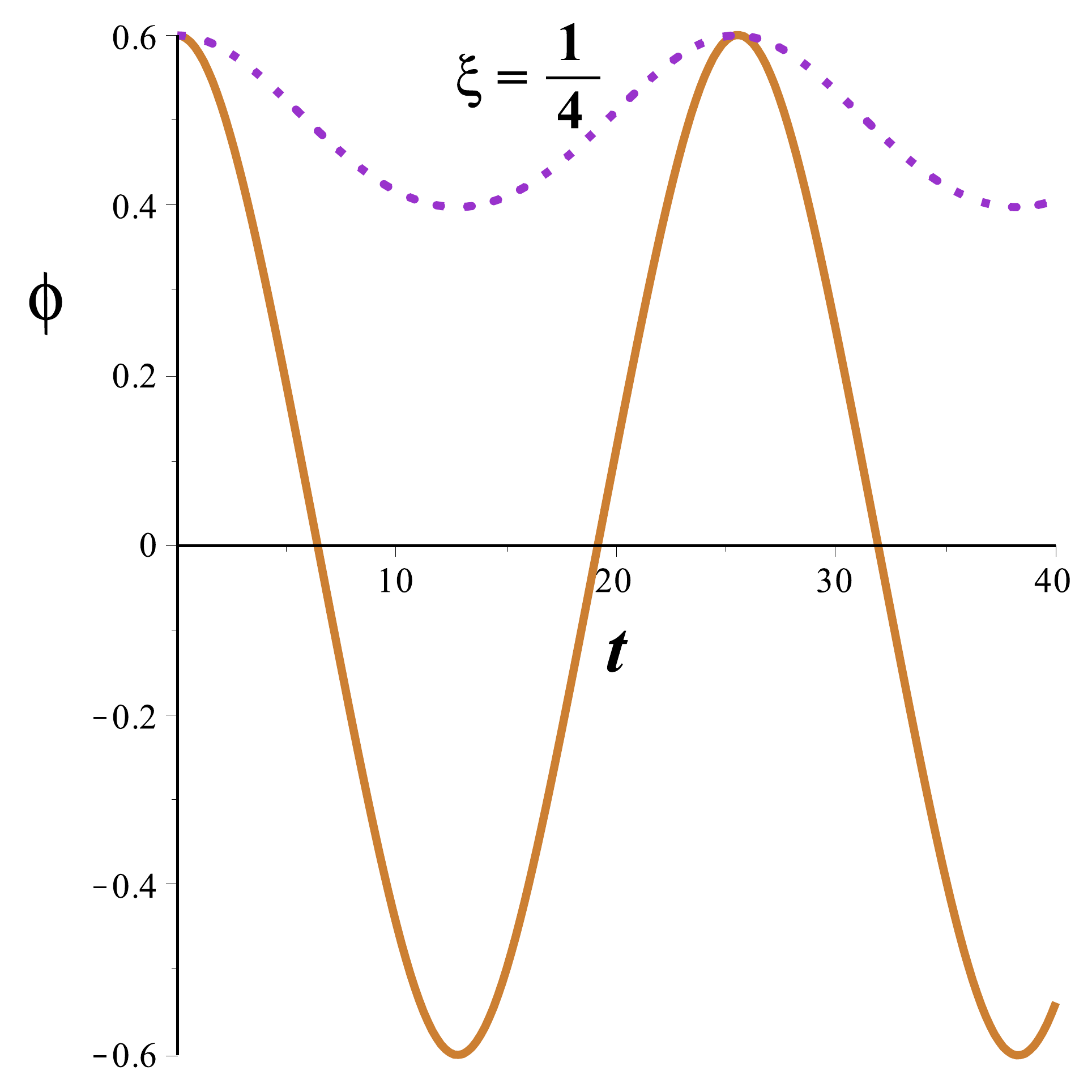} \label{fig2f}   }
\caption{\label{fig2} \emph{Emergent symmetry restoration, $\xi$ dependence.}
\small{The dotted and the solid lines represent correspondingly the bare and the perturbed solutions.
The parameters are chosen
as follows: $\mu = 0.18$, $\lambda = 0.75$, $\gamma \equiv \omega\cdot\alpha = 0.4$, $\omega = 2500$. The initial conditions are chosen as follows: $\phi(0) = 0.6$, $\dot\phi(0) = 0$.}}
\end{figure*}

Below we solve numerically the exact \eqref{geneq} and the effective \eqref{final} equations of motion for various 
values of the parameters $\gamma$ and $\xi$. The slowly varying background metric is chosen to be flat 
$\bar g_{\mu\nu} = \eta_{\mu\nu}$, and the perturbations are taken of the form \eqref{altern}. Since the metric $g_{\mu\nu}$ depends just on time, for the sake of simplicity we consider the field $\phi$ to be independent on the
spatial coordinates, i.e. $\phi = \phi(t)$. In such a setup the exact and the effective
equations correspondingly read:
\begin{figure*}[p]
\centering
\includegraphics[scale = 0.6]{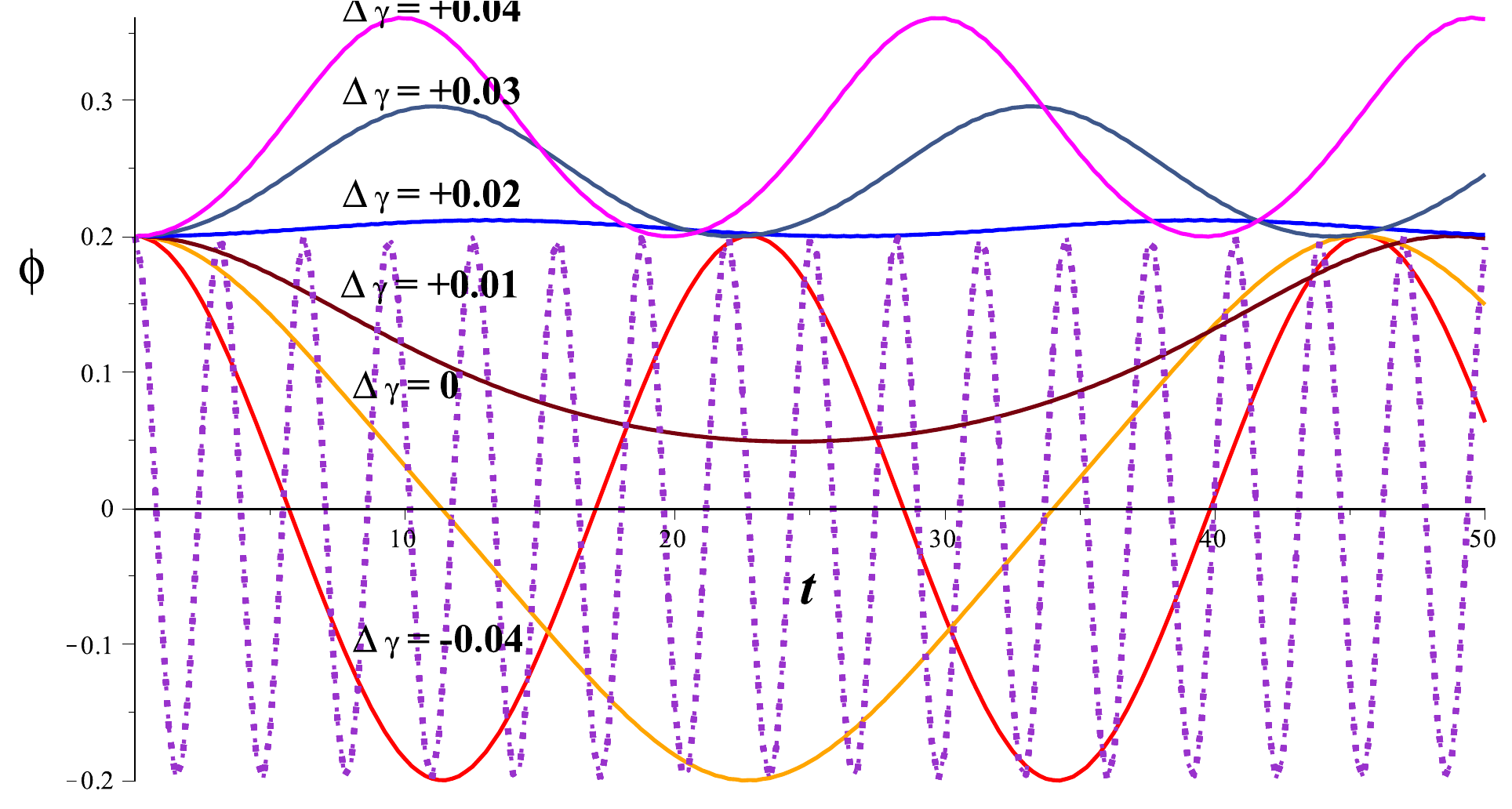}
\caption{\label{fig3} \emph{ Emergent spontaneous symmetry breaking,  $\gamma$ dependence.} 
\small{
The solid lines
represent $\phi(t)$ for various choices of the parameter $\gamma$ in the vicinity
of its critical value $\gamma_0$. The dotted line represents  the  solution of the bare equation.
Above each solid curve the deviation $\Delta \gamma\equiv \gamma-\gamma_0$ 
from the critical value is specified. 
 The parameters are chosen as follows: $\xi = +\frac{1}{12}$, $\mu = 2\ii$, hence $\gamma_0\simeq 5.657$;
 $\lambda = 4$,
  $\omega = 2500$; for each curve  $\alpha = (\gamma_0 +\Delta\gamma)/\omega $. The initial conditions are chosen as follows: $\phi(0) = 0.2$, $\dot\phi(0) = 0$.} }
\end{figure*}
\begin{figure*}[p]
\centering
\subfigure[]{\includegraphics[scale = 0.28]{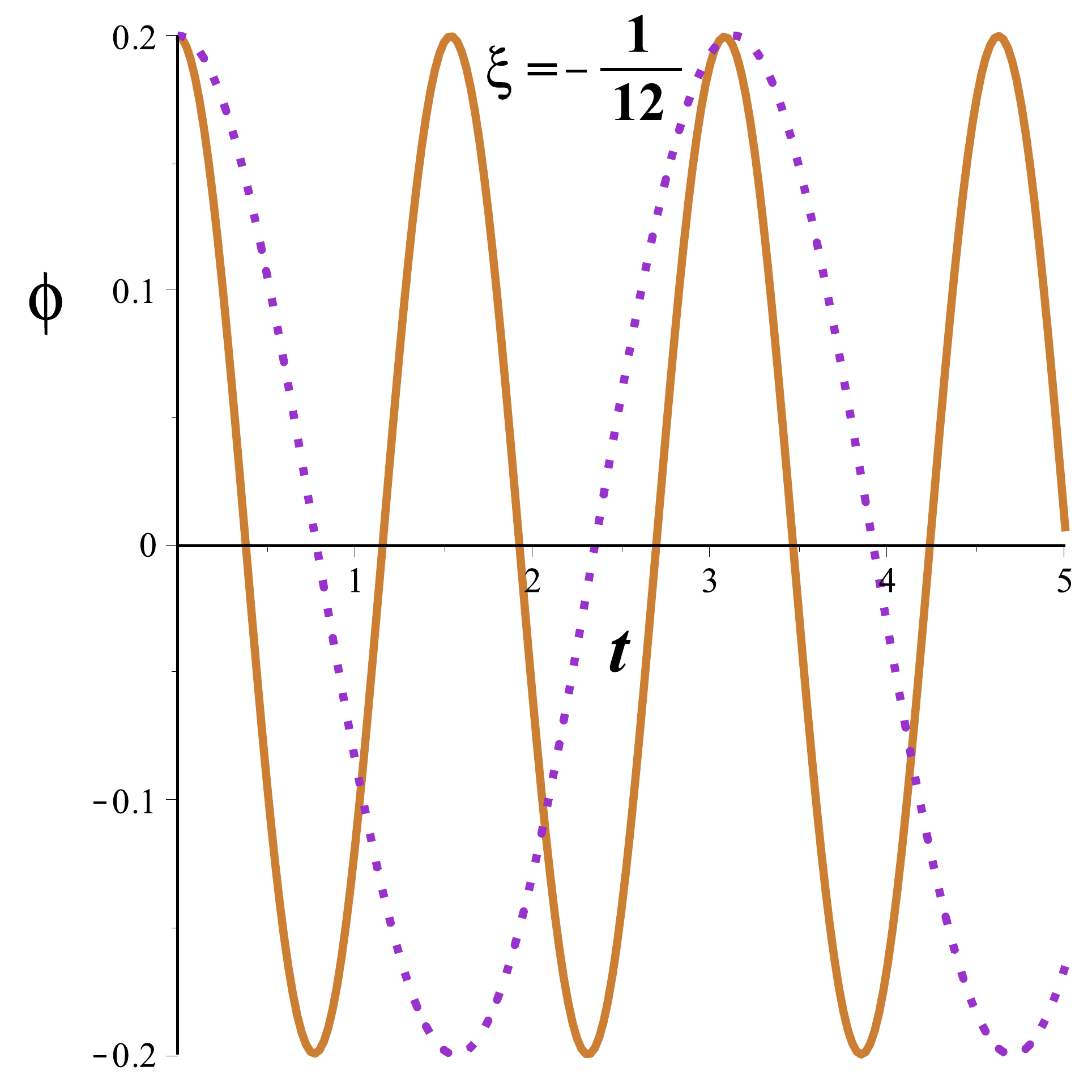} \label{fig4a} }
\subfigure[]{\includegraphics[scale = 0.28]{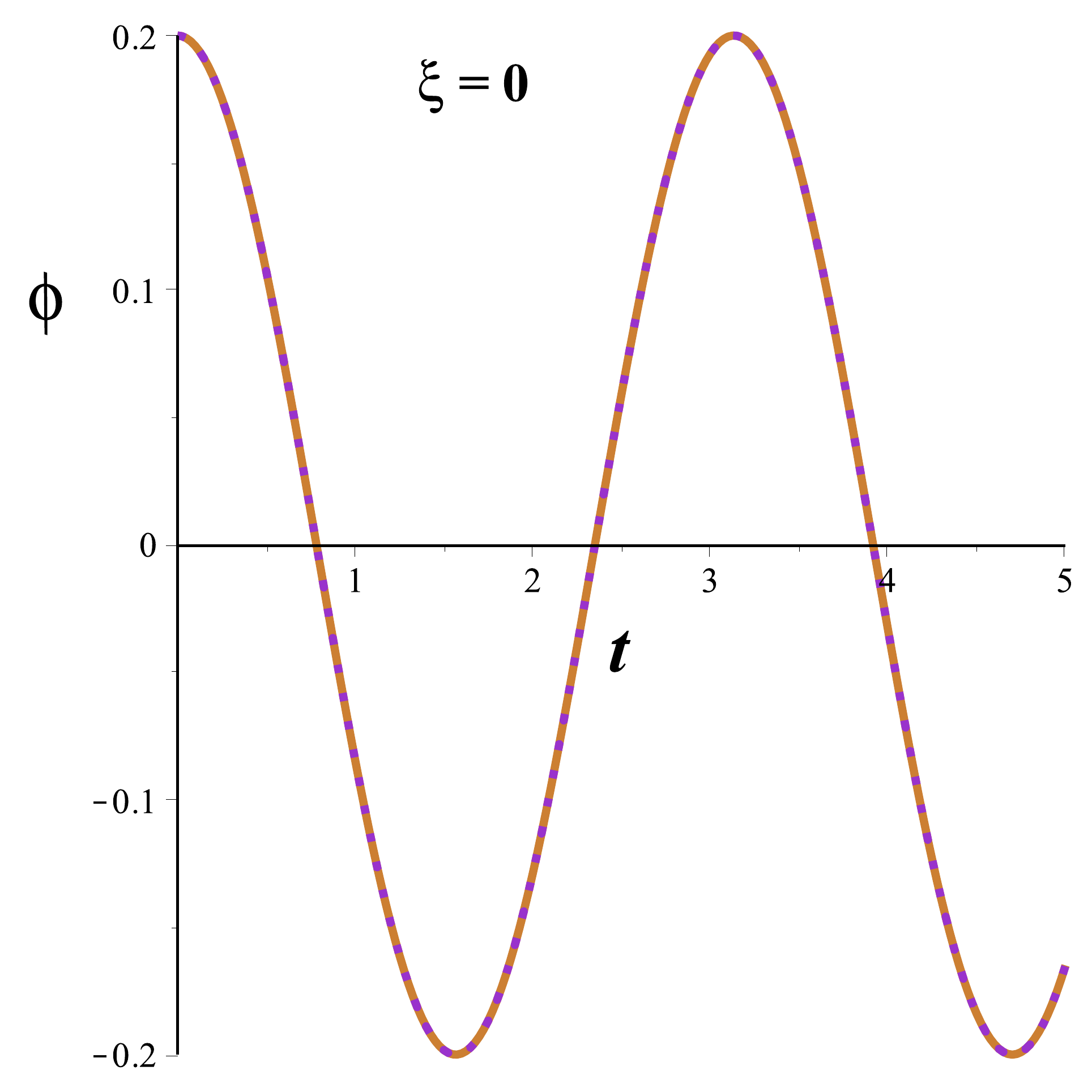} \label{fig4b} }
\subfigure[]{\includegraphics[scale = 0.28]{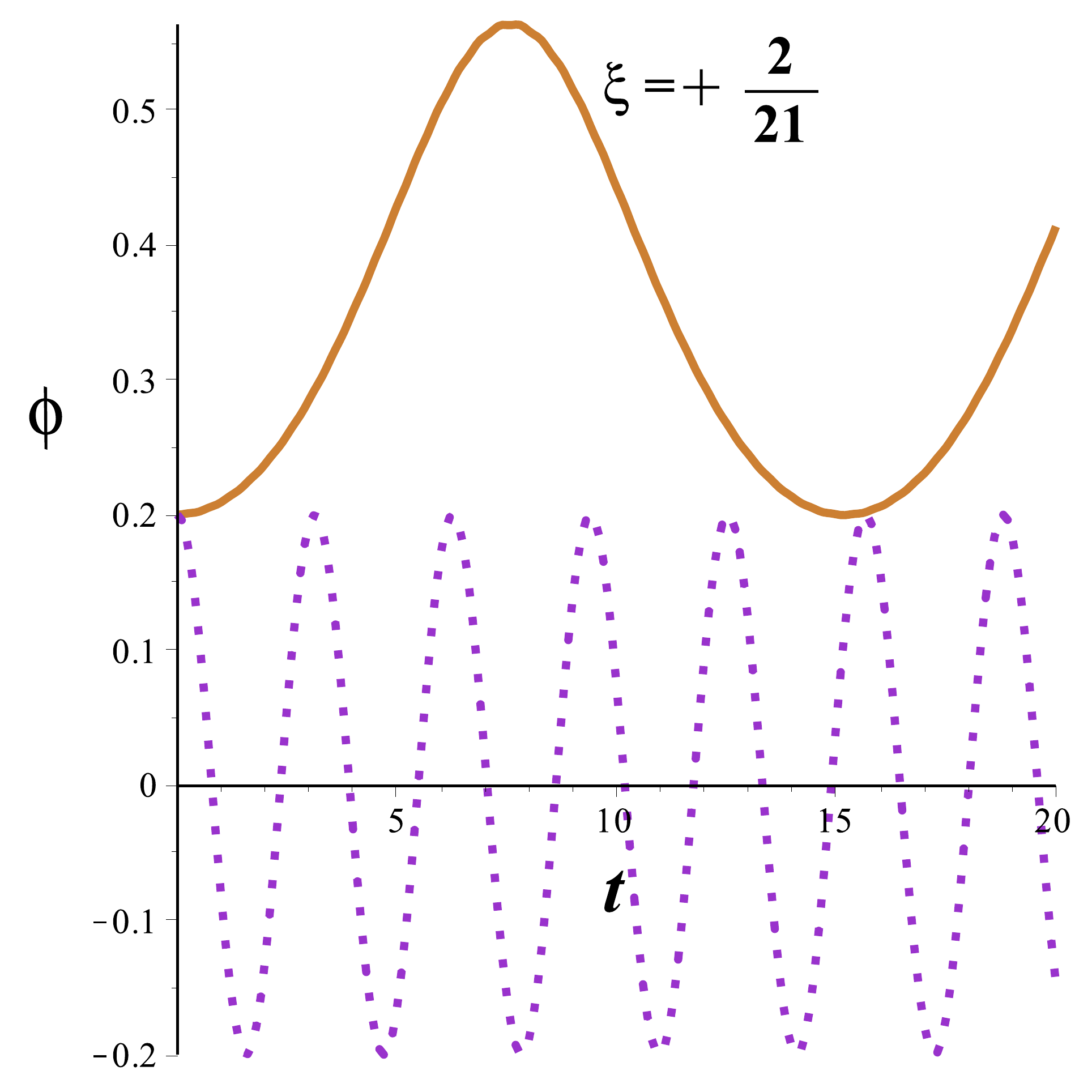} \label{fig4c} }
\subfigure[]{\includegraphics[scale = 0.28]{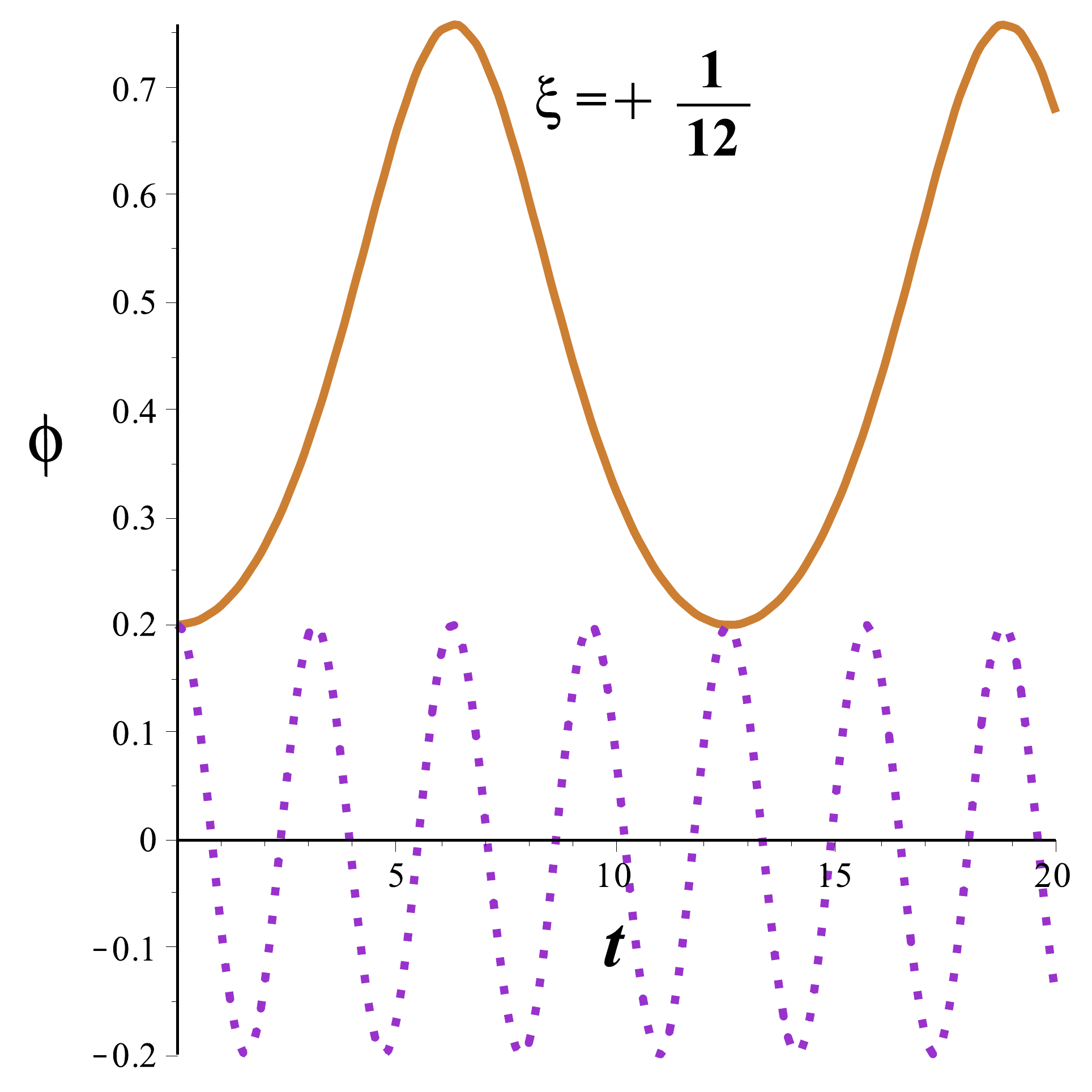} \label{fig4d} }
\subfigure[]{\includegraphics[scale = 0.28]{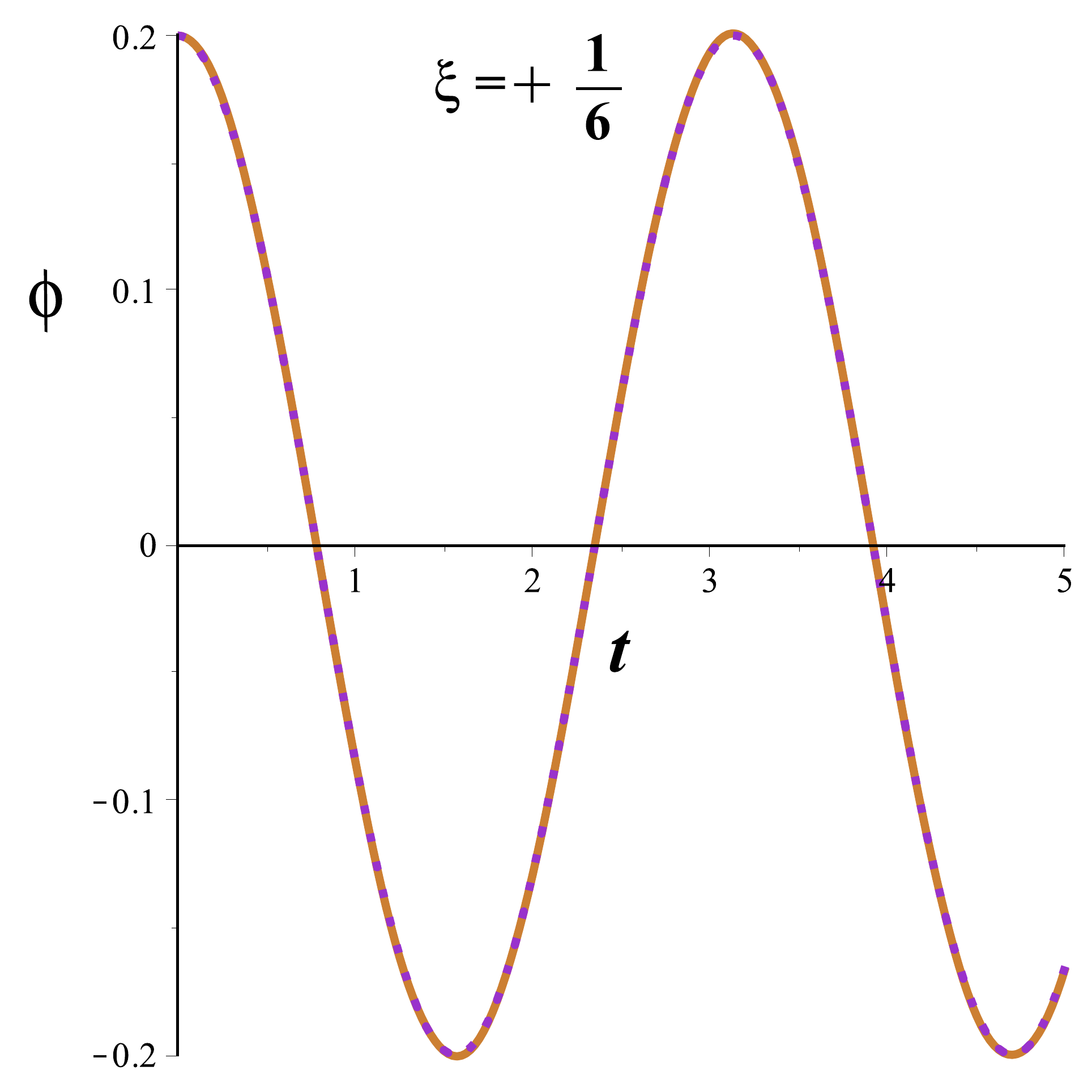} \label{fig4e} }
\subfigure[]{\includegraphics[scale = 0.28]{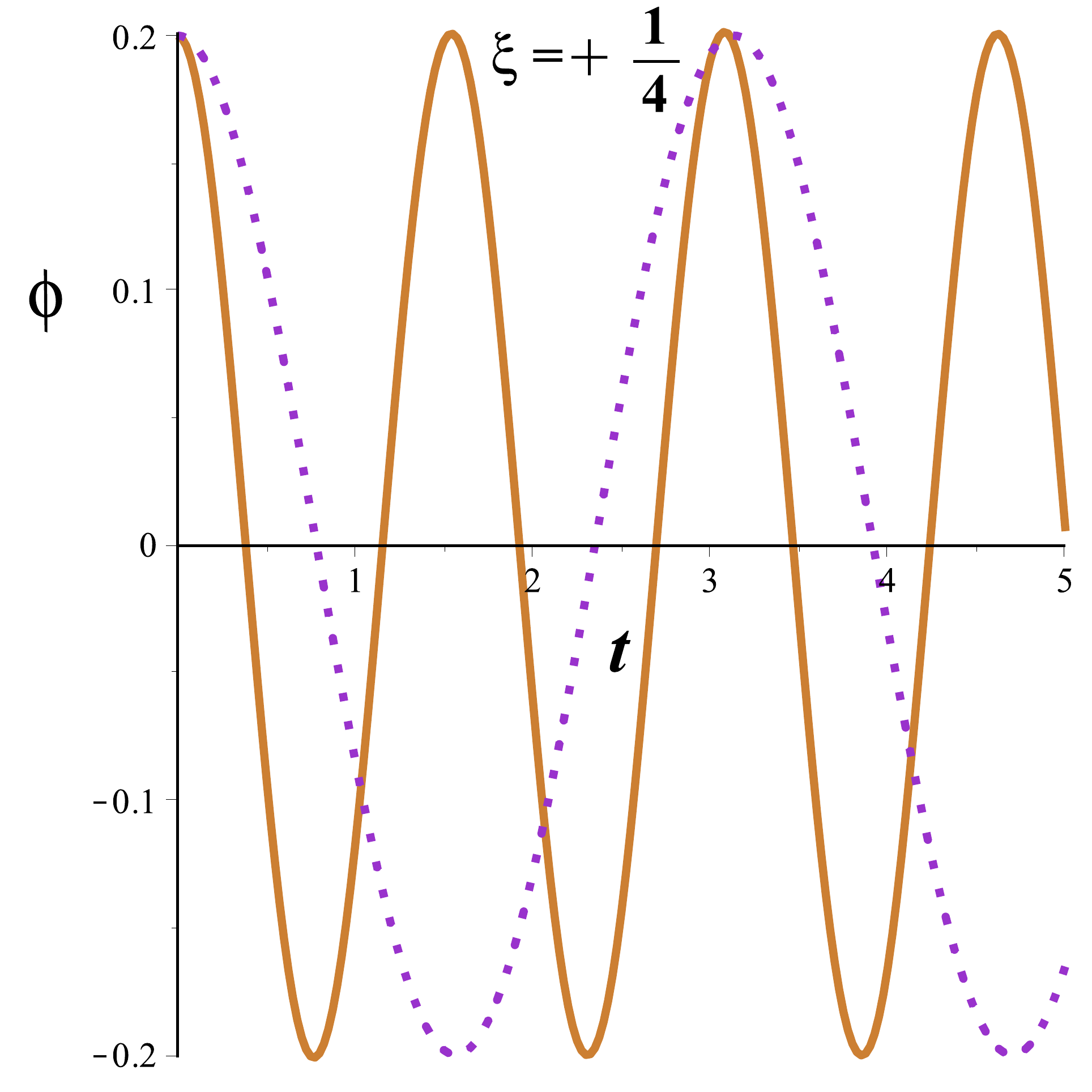} \label{fig4f}  }
\caption{\label{fig4} \emph{Emergent spontaneous symmetry breaking, $\xi$ dependence.} 
\small{
The dotted and the solid lines represent correspondingly the bare and the perturbed solutions.
The parameters are chosen as follows:
 $\mu = 2\ii$, $\lambda=4$, $\gamma \equiv \omega\cdot\alpha = 5.8$, $\omega = 2500$. The initial conditions are chosen as follows: $\phi(0) = 0.2$, $\dot\phi(0) = 0$.}}
\end{figure*}
\begin{figure*}[t]
\centering
\subfigure[]{\includegraphics[scale = 0.4]{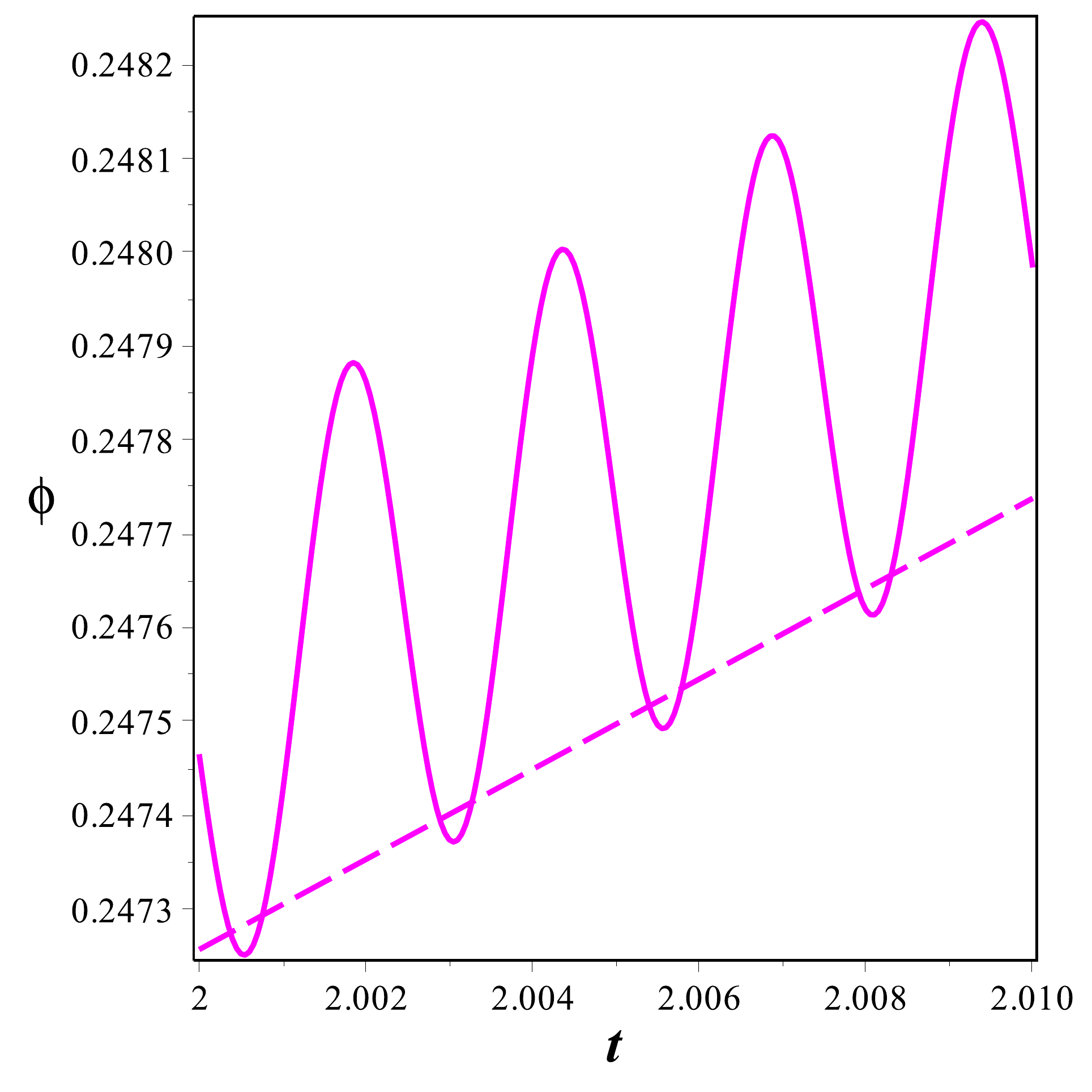} \label{fig5a}}
\subfigure[]{\includegraphics[scale = 0.4]{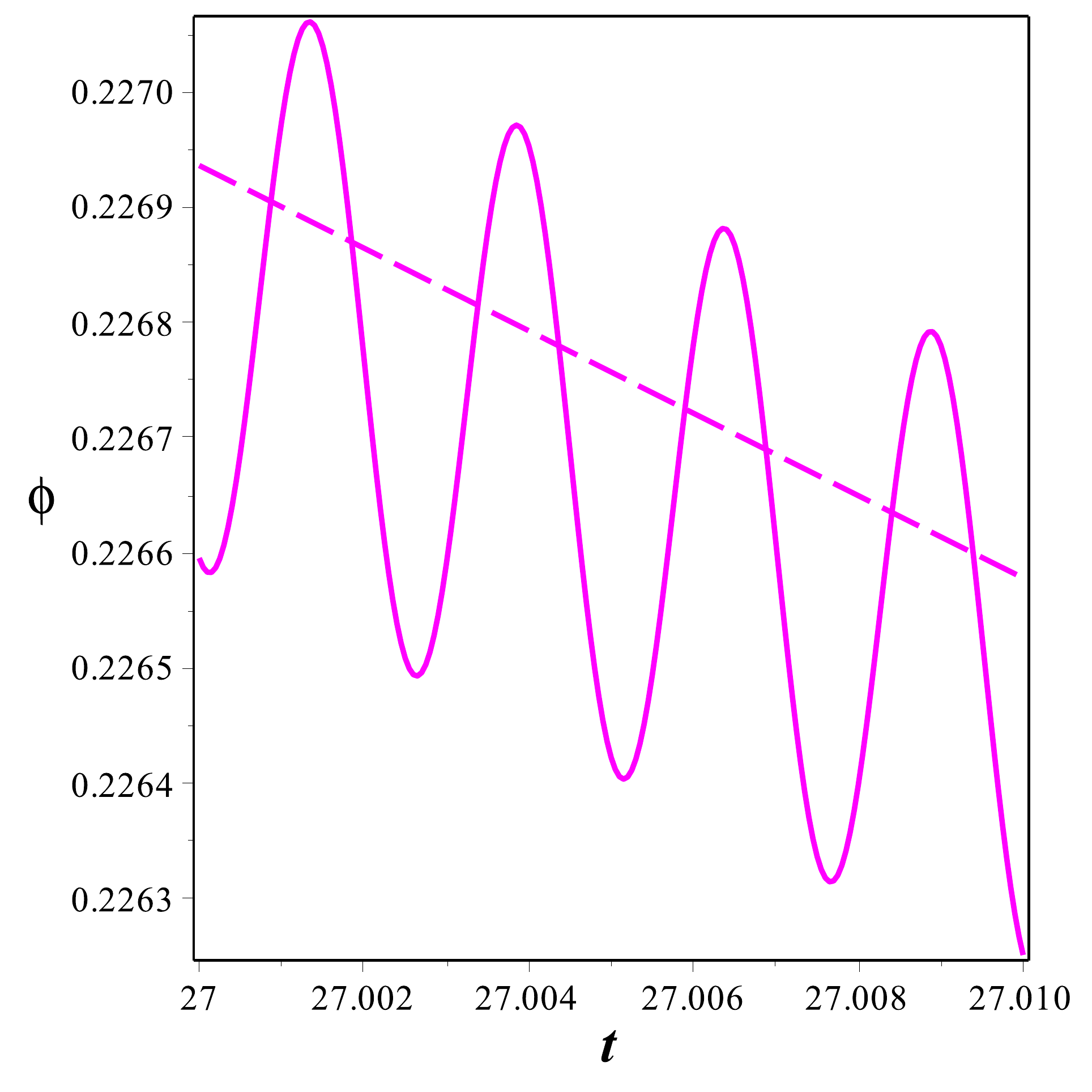} \label{fig5b}}
\caption{\label{fig5} \emph{Exact and approximate solutions: high resolution.} 
\small{
The dash and the solid lines represent  solutions of the effective  and the exact equations respectively.
The parameters are chosen as follows:  $\mu = 2\ii$, $\lambda=4$, $\gamma \equiv \omega\cdot\alpha = 4 \sqrt{2} +0.1$,  $\omega = 2500$, $\xi = 1/12$. The initial conditions are chosen as follows: $\phi(0) = 0.2$, $\dot\phi(0) = 0$. }}
\end{figure*} 

\bea
&& \ddot{\phi}  -3  \,\gamma\,\sin \left( \omega\,t
 \right)\,\dot\phi+\left(  -6\,\xi\, \omega \gamma\,\cos \left( 
\omega\,t \right) +6\,\xi\, \gamma^2 \right.\nonumber\\
&&\left. - 6\,\xi\,  {\gamma}^{2}\cos \left( 
2\,\omega\,t \right) - {\mu}^{2}\right)\phi  + \frac{1}{3!}
\,\lambda\,  \phi^{3}=0, \label{eqnumexact}
\eea
and
\be
\ddot{\phi}  - \underbrace{\left( {\mu}^{2}+ 3\gamma^2\xi (1-6\xi) \right)}_{\mu_{\rm eff}^2}\phi  + \frac{1}{3!}
\,\lambda\,  \phi^{3}=0, \label{eqnumeff}
\ee
where $\gamma\equiv\alpha\cdot\omega$. 
On the figures Fig.~\ref{fig1}, Fig.~\ref{fig2}, Fig.~\ref{fig3} and Fig.~\ref{fig4} numerical solutions of the exact  
 and the effective   equations are undistinguishable in agreement with the fact that the difference
between the exact and the effective solutions is of the order of $\omega^{-1} \sim 10^{-3}$ for our choice of $\omega$. For a short distance behavior see Fig.~\ref{fig5}.
Throughout this section\footnote{c.f. Eq.~\eqref{cond1} and Eq.~\eqref{cond2}}
 $\gamma_0\equiv\frac{|\mu|}{ \sqrt{|3\xi(6\xi - 1)|}}$ stands for  the critical value of the parameter $\gamma$. 

Let us consider the system which is spontaneously broken\footnote{ i.e. $\mu^2  >0 $} at $\omega = 0$.
 Varying the parameter $\gamma\equiv \alpha\cdot\omega$ above and below its critical value we solve numerically the equations
 \eqref{eqnumexact}, \eqref{eqnumeff}; corresponding solutions are presented
on Fig.~\ref{fig1}.  As one would have expected for small negative deviations
$\Delta\gamma$  of $\gamma$ from the critical value $\gamma_0$ the effect of rapid oscillations results in a shift of the frequency
of slow oscillations and a shift of the (nonzero)
equilibrium point. When $\Delta\gamma \geq 0$ the solutions with high precision 
are described by slow oscillations around the symmetric ground state $\phi = 0$, so effectively at the scales of length much higher than
$\omega^{-1}$ the system behaves as \emph{non} spontaneously broken. 

Corresponding dependence on  the non minimal scalar-tensor coupling $\xi$ is presented on Fig.~\ref{fig2}. 
Inside the interval $0<\xi<1/6$ the only effect is a shift of the frequency of small oscillations around the shifted equilibrium
point $\phi\neq 0$, see Fig.~\ref{fig2c} and Fig.~\ref{fig2d}. At  $\xi = 0$ and $\xi = \frac{1}{6}$
the effect is absent: the bare ($\omega = 0$) and the perturbed curves (up to small deviations $\sim \omega^{-1}$) coincide, Fig.~\ref{fig2b}
and Fig.~\ref{fig2e}.
Finally at $\xi < 0 $ and at $\xi>\frac{1}{6}$ for a high enough $\gamma$ the emergent symmetry restoration takes place, see Fig.~\ref{fig2a} and
Fig.~\ref{fig2f}.

Now let us elaborate in a similar fashion the  spontaneously unbroken\footnote{i.e. $\mu^2  = -m^2  \leq 0 $} system at $\omega = 0$. For various choices of the parameter $\gamma\equiv \alpha\cdot\omega$ above and below its critical value numerical solutions of \eqref{eqnumexact}, \eqref{eqnumeff} are presented
on Fig.\ref{fig3}.  Again for small negative deviations
$\Delta\gamma$  of $\gamma$ from the critical value $\gamma_0$ the effect of rapid oscillations results in a  shift of  the frequency of slow oscillations around the equilibrium point $\phi = 0$. When $\Delta\gamma \geq 0$ the solutions with high precision are described by slow oscillations around the ground  state at $\phi \neq 0$, which corresponds to spontaneously broken symmetry. Thus we conclude that effectively at the scales of length much higher  than
$\omega^{-1}$ the system behaves as  spontaneously broken.

On Fig.~\ref{fig4} we illustrate the dependence on the parameter $\xi$  for this situation. At $\xi\leq 0$ and $\xi \geq 1/6$
the only effect is a change of the frequency of slow oscillations around the symmetric ground state $\phi = 0$, see Fig.~\ref{fig4a} and Fig.~\ref{fig4f},
in particular at  $\xi = 0$ and $\xi = 1/6$ at the scales much higher than $\omega^{-1}$ there is no any effect
at all, see Fig.~\ref{fig4b} and Fig.~\ref{fig4e} . Inside the interval $0<\xi<1/6$ for sufficiently high values of $\gamma$ one has the phase transition
 - emergent spontaneous symmetry breaking, see Fig.~\ref{fig4c} and Fig.~\ref{fig4d}.

In conclusion we present typical solutions of the exact \eqref{eqnumexact} and the effective \eqref{eqnumeff} equations  
 at \emph{small} scales, which are comparable with $\omega^{-1}$. On Fig.~\ref{fig5a}
and Fig.~\ref{fig5b} the small scale behavior is presented for various time intervals.  One can see that the the exact solution  oscillates with the 
frequency $\omega$, while the difference between the two\footnote{Since deriving the effective equation we passed to the limit $\omega\rightarrow\infty$,
 we neglected by terms of the order of $\mathcal{O}\left(\omega^{-1}\right)$, thus the mean value
of the precise solution and the effective solution may slightly differ, however this difference if of the order of $\mathcal{O}\left(\omega^{-1}\right)$.} is indeed of the order of $\omega^{-1}\sim 10^{-3}$.

\begin{acknowledgements}
The author is grateful to  F. Lizzi, O. Novikov  and D. Vassilevich for reading  the manuscript and  constructive comments. This work is supported  by the FAPESP process 2015/05120-0. 
\end{acknowledgements}


\begin{thebibliography}{9}

\bibitem{Power:1998wh}
  W.~L.~Power and I.~C.~Percival,
  ``Decoherence of quantum wavepackets due to interaction with conformal space-time fluctuations,''
  Proc.\ Roy.\ Soc.\ Lond.\ A {\bf 456} (2000) 955
  doi:10.1098/rspa.2000.0544
  [quant-ph/9811059].
 \bibitem{Wang:2006vh}
  C.~H.-T.~Wang, R.~Bingham and J.~T.~Mendonca,
  ``Quantum gravitational decoherence of matter waves,''
  Class.\ Quant.\ Grav.\  {\bf 23} (2006) L59
  doi:10.1088/0264-9381/23/18/L01
  
\bibitem{Durrer:1995mz}
  R.~Durrer and J.~Laukenmann,
  ``The Oscillating universe: An Alternative to inflation,''
  Class.\ Quant.\ Grav.\  {\bf 13} (1996) 1069
  doi:10.1088/0264-9381/13/5/021
  
  
\bibitem{Collins:2001ic}
  H.~Collins and B.~Holdom,
  ``Matter in a warped and oscillating background,''
  Phys.\ Rev.\ D {\bf 65} (2002) 124014
  doi:10.1103/PhysRevD.65.124014
  
\bibitem{Collins:2001mz}
  H.~Collins and B.~Holdom,
  ``Decay of the cosmological constant through an oscillating metric,''
  JHEP {\bf 0210} (2002) 052
  doi:10.1088/1126-6708/2002/10/052
  
\bibitem{Collins:2001ni}
  H.~Collins and B.~Holdom,
  ``The Randall-Sundrum scenario with an extra warped dimension,''
  Phys.\ Rev.\ D {\bf 64} (2001) 064003
  doi:10.1103/PhysRevD.64.064003
  
\bibitem{Hwang:1993cv}
  J.~C.~Hwang,
  ``Curved space quantum scalar field theory with accompanying metric fluctuations,''
  Phys.\ Rev.\ D {\bf 48} (1993) 3544.
  doi:10.1103/PhysRevD.48.3544  

%
\bibitem{Mangano:2015pha}
  G.~Mangano, F.~Lizzi and A.~Porzio,
  ``Inconstant Planck's constant,''
  Int.\ J.\ Mod.\ Phys.\ A {\bf 30} (2015) 34,  1550209
  doi:10.1142/S0217751X15502097

\bibitem{deCesare:2016dnp}
  M.~de Cesare, F.~Lizzi and M.~Sakellariadou,
  ``Effective cosmological constant induced by stochastic fluctuations of Newton's constant,'' 
   arXiv:1603.04170 [gr-qc].
 
\bibitem{Kapitza1} P. L. Kapitza, "Dynamic stability of a pendulum when its point of suspension vibrates", Soviet Phys. JETP 21, 588–592 (1951); 
\bibitem{Kapitza2}
P. L. Kapitza, "Pendulum with a vibrating suspension" Usp. Fiz. Nauk, 44, 7-15 (1951).

\bibitem{Vilenkin:1982wt}
  A.~Vilenkin and L.~H.~Ford,
  ``Gravitational Effects upon Cosmological Phase Transitions,''
  Phys.\ Rev.\ D {\bf 26} (1982) 1231.
  doi:10.1103/PhysRevD.26.1231
  


\end{thebibliography}
\end{document}